# A Strategic Decision Framework for Enterprise LLM Adoption

by Michael Trusov, Minha Hwang, Zainab Jamal, Swarup Chandra

## Summary

Organizations are rapidly adopting Large Language Models (LLMs) to transform their operations, yet they lack clear guidance on key decisions for adoption and implementation. While LLMs offer powerful capabilities in content generation, assisted coding, and process automation, businesses face critical challenges in data security, LLM solution development approach, infrastructure requirements, and deployment strategies. Healthcare providers must protect patient data while leveraging LLMs for medical analysis, financial institutions need to balance automated customer service with regulatory compliance, and software companies seek to enhance development productivity while maintaining code security.

This article presents a systematic six-step decision framework for LLM adoption, helping organizations navigate from initial application selection to final deployment. Based on extensive interviews and analysis of successful and failed implementations, our framework provides practical guidance for business leaders to align technological capabilities with business objectives. Through key decision points and real-world examples from both B2B and B2C contexts, organizations can make informed decisions about LLM adoption while ensuring secure and efficient integration across various use cases, from customer service automation to content creation and advanced analytics.

**The Proposed Decision Framework**

As organizations increasingly adopt LLMs, they face crucial implementation decisions that determine success. This proposed framework examines key decision points in the LLM adoption journey, from (1) business application identification, (2) LLM solution building approach between on-premises building and third-party solution, (3) LLM model adaptation approaches, (4) data strategies for model refining (finetuning), (5) LLM performance evaluation, and (6) LLM deployment options. Figure. 1 shows our proposed LLM adoption decision framework. We discuss this step-by-step in the following sections.

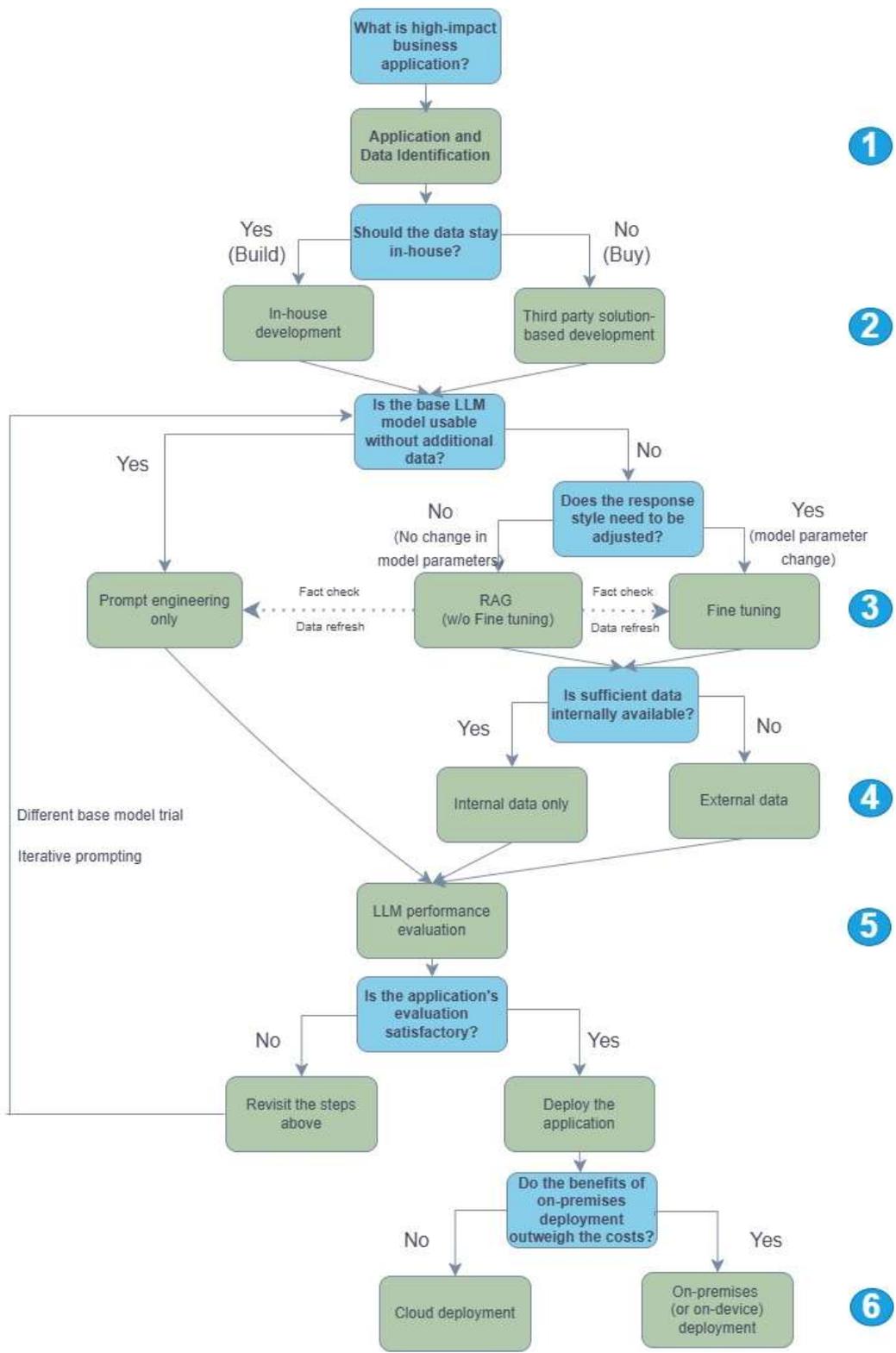

**Fig 1. LLM Adoption Decision Framework – Six Steps**

**The Six-Step Decision Framework**

**Step 1: High Impact Business Application Identification**

Organizations must first identify high-impact business applications for LLM integration. Focus areas include content creation, summarization and personalization, software development, customer service automation, data analysis, and conversational assistants [1]. This step requires evaluating data availability, quality, and accessibility while considering potential business value and implementation feasibility. Ideally cost vs. benefit analyses need to be conducted to assess the potential impacts from the adoption.

*Content Creation, Summarization and Personalization*

LLMs excel at generating and personalizing content across marketing and customer engagement channels. These models can rapidly create tailored advertising materials, website content, and marketing collateral for specific audience segments based on demographics, behavior patterns, and customer status. The technology enables real-time content delivery and highly personalized customer experiences by drawing from existing content repositories and customer data. It can also enhance relevance by incorporating up-to-date information through techniques like Retrieval Augmented Generation (RAG), which supplements AI-generated responses with relevant, real-time data from external sources. As an example, Typeface AI demonstrates this through their personalized ad copy generation platform, which analyzes customer data and brand guidelines to produce targeted marketing contents. One company leveraging this approach is a Fortune 500 automaker who utilized Typeface's AI ad generator to create targeted paid advertisements for their year-end sales event [2]. This approach enabled the creation of four times more personalized ad variations across eight target audience segments in 57% less time, all while maintaining brand consistency [2]. Similarly, Jasper AI supported BestPlaces to revitalize their content by automated creation of over 80K unique header paragraphs describing different locations [3]. This led to an 800% surge in website traffic [3].

In B2B applications, LLMs enhance market intelligence by analyzing CRM data and customer interactions, empowering sales teams with deeper insights for customer engagement. The models can process multiple data sources simultaneously, providing comprehensive summaries and actionable intelligence while maintaining content relevance and accuracy through continuous data access. According to a McKinsey report, a large European telco developed a gen-AI-powered dashboard to call center managers and sellers that analyzed customer service call scripts, scored conversation performance, identified skill improvement opportunities, and created dedicated coaching programs for sellers [4]. This led the telco to a 20 – 30% improvement in customer satisfaction [4].

### *Productivity Enhancement in Software Development*

LLMs are transforming software development by converting natural language requirements into functional codes. They enhance developer productivity through automated code generation, intelligent completions, assisted debugging and unit testing, and streamlined documentation. This allows developers to focus on complex architectural decisions and business logic. For example, [GitHub Copilot](#) has demonstrated a 55% reduction in coding time of developers [5]. One AI-focused startup reported cutting their development time for a minimum viable product (MVP) by 40% after integrating Cursor AI into their workflow [6]. These AI-powered tools seamlessly integrate with popular integrated development environments (IDEs) like Visual Studio Code, enhancing the natural workflow of software development.

Organizations implementing LLMs must address key challenges including intellectual property protection, code security, and privacy standards. By carefully balancing automation benefits with security considerations, organizations can significantly enhance their software development capabilities while effectively managing risks.

### *Insight Generation from Business Intelligence Enhancement*

LLMs are revolutionizing how organizations extract insights from business data across Marketing, Sales, and Finance functions. Moving beyond traditional dashboard interfaces, these models convert complex data queries into natural language interactions and transform analytical results into clear business insights. Technology enhances existing BI applications by maintaining context across user interactions and personalizing outputs for different roles - from analysts to executives - while learning from both individual user patterns and domain knowledge.

This new approach to business intelligence combines the analytical power of traditional BI tools with the intuitive interface of conversational AI, making data analysis more accessible and efficient across organizational levels.

### *Conversational AI and Enhanced Search*

LLMs are transforming both human-AI interactions and information retrieval capabilities. Modern conversational assistants (like ChatGPT, Copilot, Claude, and Gemini) provide contextual, real-time responses that enhance both automated customer service and human agent effectiveness. For example, Microsoft's implementation of Copilot in one area of commercial support business resulted in a 12% reduction in average handle time for low-severity chat cases [7]. LLMs also power next-generation search engines such as Google's AI Overviews, Microsoft Bing's Deep Search, and Perplexity - by processing complex queries and synthesizing information from multiple sources into comprehensive answers [8, 9].

The landscape extends beyond major providers to include rapidly evolving open-source LLMs, which offer cost-effective alternatives for specialized applications. DeepSeek-R1 achieved comparable reasoning capabilities to OpenAI's models at just 3-5% of the cost ($5.68 million budget) [10], while Mistral AI's open-source 7B model demonstrated enterprise-grade performance in specific tasks while operating at a fraction of the computational cost of larger models [11]. Notably, it outperforms Meta's Llama 2 13B model on all benchmarks and rivals the performance of the Llama1 34B model in many areas [11].

**Step 2: In-house vs. Third-party Development – Build or Buy**

When implementing LLM applications, organizations must strategically choose between in-house solutions and third-party solutions that fundamentally impact their data independence and autonomy - whether to keep their data in-house (on-premises) or share it with third parties. This crucial build or buy decision affects every aspect of implementation - model selection, model adaptations, user interactions and application deployment, monitoring and maintenance, ultimately impacting security, privacy, and compliance requirements. Organizations typically choose between two main approaches: building in-house solutions or buying third-party solutions.

In-house LLM developments offer enhanced security, extensive customization capabilities, and consistent performance. Organizations can utilize open-source models like Llama 2 and Mistral through platforms like Hugging Face, though these models typically have fewer parameters than closed-source alternatives. Open-source LLMs like Llama, Mistral or DeepSeek, Tulu 3 are key enablers for the in-house developments [10, 11]. However, this approach requires significant infrastructure investments and specialized technical expertise for implementation and maintenance.

For organizations seeking a more lightweight and efficient alternative, Small Language Models (SLMs) provide a viable option. These compact AI models are optimized for local processing, delivering faster response times while reducing reliance on cloud-based infrastructure. On-device LLM development based on SLMs has emerged as a crucial solution for applications requiring minimal latency and enhanced privacy protection. The application areas are games and portable smart devices. NVIDIA introduced an on-device SLM named Nemotron-4 4B Instruct, designed to enhance the conversational abilities of game characters [12]. This model is optimized to run efficiently on local devices by reducing its size and computational requirements, ensuring faster performance and smoother on-device processing. [12]. Memory usage has been minimized to 2 GB with faster execution compared to LLMs [12]. It is showcased in the game *Mecha BREAK* by Amazing Seasun Games [12]. Stability AI released Stable LM 3B, a compact 3-billion-parameter language model designed for portable digital devices like smartphones and laptops [13]. Despite its smaller size, Stable LM 3B outperforms previous state-of-the-art models of similar scale and even some of the best open-source language models at the 7B parameter scale [13]. These developments demonstrate how SLMs can deliver powerful capabilities within the constraints of edge devices while preserving user privacy and being environmentally more friendly [13].

Third-party LLM solutions, while offering faster developments and requiring minimal infrastructure changes, come with their own trade-offs. These solutions need less technical expertise and lower initial investment, but raise concerns about data privacy and security, limited control over the model, and potential service disruptions. While initially cost-effective, long-term expenses may exceed those of on-premises solutions. Additionally, though third-party solution providers offer some customization options, they generally provide less flexibility than in-house developments. Recent research by Kai Greshake and colleagues as well as research by Microsoft have shown that prompt-tuned LLMs are vulnerable to prompt injection attacks, where malicious inputs can manipulate model outputs [14, 15].

The choice between these options depends on an organization's available resources, customization requirements, and performance expectations. Making this assessment early in the implementation process is essential for selecting the most suitable LLM solution.

**Step 3: LLM Model Adaptation Approaches**

When selecting an LLM solution, organizations must consider both the application domain and adaptation methods. The choice between third-party and on-premises LLM solutions significantly impacts feasible model adaptation options. Third-party LLMs offer general-purpose capabilities but limit customization to prompt engineering or prompt engineering grounded with public search browsing data. On-premises solutions provide more flexibility, allowing prompt engineering, prompt engineering enhanced by RAG with private data, and model fine-tuning within their ecosystems. On-premises solutions using open-source LLMs offer maximum customization but require substantial technical expertise and resources.

For model adaptation, organizations can choose between prompt engineering, retrieval-augmented generation (RAG), and fine-tuning approaches [16]. Prompt engineering, applicable across all LLM types, involves designing task-specific inputs (i.e., prompts) to guide model responses. While this method requires no parameter updates, its effectiveness depends on the alignment between the task and the available model's style (e.g., chat, reasoning, etc.) Given that most closed third-party models are instruction tuned for chat responses, a prompt engineering approach tends to struggle with non-chat applications. In addition, it is easy to get a quick demo from prompt engineering, but it is very common to struggle to scale these solution to production. Even when prompt engineering is a viable solution, the developments will include trials with many different base models (e.g. OpenAI GPT-4o, o1, or o1-mini) and iterative prompting trials. Tools like Prompt Source can help create structured prompts, though manual prompt design often involves trial-and-error optimization. ~ 30 – 50 prompt iterations are quite common in practice.

RAG enhances prompt engineering by incorporating relevant external knowledge during inference [16]. This approach is particularly effective when dealing with domain-specific tasks or when current, accurate information is crucial [16]. RAG also helps to provide references for fact checking for the responses. Unlike fine-tuning, RAG doesn't modify model parameters but instead augments the context with retrieved information, making it more suitable for tasks requiring access to up-to-date or specialized knowledge [16]. RAG is preferred when the task requires factual accuracy, domain expertise, or when data frequently changes. A well-known example is ChatGPT with grounding from public browsing, which addresses hallucination, current data retrieval beyond model training cut-off date, and provision of URL links for fact checking. It is important to note that RAG and fine tuning are not mutually exclusive. It is possible to apply RAG to fine tuned models.

Fine-tuning becomes necessary for complex NLP tasks where prompt engineering proves challenging. This method modifies model parameters using task-specific training data, requiring more computational resources but potentially yielding better performance. Parameter-efficient tuning methods can help manage resource constraints, though practitioners must still carefully consider factors like batch size, learning rate, and training epochs. Fine-tuning is particularly valuable when tasks significantly deviate from the model's pre-training objectives or require a significant change in styles in model response

[16]. Simply put, fine tuning is required for change in model response style. In contrast, RAG is helpful to address factual accuracy and frequent data changes.

The choice between these approaches depends on factors including task complexity, resource availability, data sensitivity, factual accuracy, data change frequency, and performance requirements. Organizations should evaluate these factors alongside their technical capabilities and infrastructure constraints to determine the most suitable model adaptation approaches.

**Step 4: Data for LLM Fine-tuning - Internal Data Only vs. External Data**

Fine-tuning an LLM becomes necessary when prompt engineering fails to meet performance requirements. While training a new LLM from scratch is typically unfeasible for most organizations, fine-tuning offers a practical approach to model adaptation from widely distributed base models. The success of fine-tuning heavily depends on identifying and obtaining high-quality training data. The quality of data matters more than mere amounts of the data. RAG with private data retrieval is another possible approach.

Organizations can leverage internal data sources, such as existing databases, customer service logs, or domain-specific documents. For instance, a law firm might use its archive of legal documents, while a customer service department could utilize chat logs. When internal data is limited, data augmentation techniques can enhance the training dataset.

When internal data proves insufficient, external sources provide viable alternatives. External data include public datasets (The Pile [17], Common Crawl) from platforms like Kaggle or Hugging Face, web-scraped data (where legally permissible), data providers like ACNielsen, Circana, Kantar, or Kroger: 84.51 (POS, consumer panel, advertising media), Similarweb or Comscore (Internet), IQVIA, Veeva Crossix or PRA Health Science (healthcare), Duns and Bradstreet (company database for B2B sales), Placer.ai, Unacast, or Verset (mobile foot traffic data), Kantar: Numerator (offline competitive price intelligence data from scanned supermarket flyers), and API-accessed information (e.g., Zillow, Amazon). Additionally, organizations can utilize LLMs to generate synthetic training data. This approach offers multiple benefits: allowing for accuracy verification before fine-tuning, focusing the model on specific domains, and expanding limited datasets through controlled variations of existing data.

Successful fine-tuning requires careful evaluation of data quality, relevance, and volume from both internal and external sources. High-quality data are crucial for the success of fin-tuning. With high quality data, relatively small inputs and output pairs for the task (e.g., 1K – 80K) can deliver better than okay results from non-customized models [16] (e.g. ChatGPT,

Claude.ai.) Organizations should consider legal and ethical implications while selecting data sources and ensure alignment with their specific use cases. This is even more critical for healthcare applications due to regulations such as Health Insurance Portability and Accountability Act (HIPAA).

**Step 5: LLM Performance Evaluation - Metrics and Assessment Approaches**

Evaluating LLM performance requires careful selection of metrics aligned with business objectives and application domains. For structured outputs like numeric or categorical predictions from Machine Learning (ML), this was easier with traditional metrics such as Precision, Accuracy, and RMSE as reliable performance indicators when compared against ground truth values.

Unstructured outputs, particularly in conversational applications, require specialized evaluation approaches. Text-specific metrics like BLEU, ROUGE, and METEOR measure similarity between generated and reference texts [18]. However, these are limited beyond measuring factual truthfulness and closeness to benchmark answers. When reference texts aren't available, human evaluation becomes crucial, either through expert panels or user feedback systems.

Evaluating user satisfaction for LLM applications requires a comprehensive measurement framework: both offline and online metrics. Offline measurement with human feedback provides the best view on response quality and user satisfaction. Online measurements capture real-world usage patterns across multiple dimensions: performance(e.g., time to first token render, page loading time, latency), engagement (e.g., time spent in chats), guard-rail metrics (e.g., GPU utilization rate), debugging metrics (e.g., response category breakdown of user satisfaction), and feature-area specific metrics (e.g. Q&A quality for travel segment.) Complex LLM applications often benefit from combining quantitative metrics with qualitative assessments like user studies to ensure comprehensive user understanding and performance evaluation across all relevant dimensions.

**Step 6: LLM Deployment - Cloud vs. On-premises Deployment**

When deploying LLM solutions, organizations must choose between cloud-based and on-premises implementations, each offering distinct advantages and challenges. Cloud deployments provide cost-effectiveness upfront, easier implementation, and access to the latest models but introduce concerns about vendor lock-in, data privacy and security, and ongoing operational costs. Conversely, on-premises solutions offer greater control, reduced latency, and enhanced privacy, though they require significant upfront investment and technical expertise.

Cost considerations span both infrastructure and operational aspects. While larger models with expanded context windows offer improved capabilities, they incur substantially higher costs per token. For instance, GPT-4o mini is approximately seventeen times less expensive per token than GPT-4o, with output tokens typically costing four times as much as input tokens [19]. Organizations must carefully weigh these costs against expected benefits.

Performance factors, particularly latency, significantly impact deployment decisions. While input processing can occur in parallel, sequential output generation affects response times. Cloud-based solutions may introduce network latency, which can be mitigated through strategies like output streaming and response caching. Local deployment can reduce these delays but requires robust infrastructure management.

On-device deployments offer another alternative, especially when robust SLMs equipped with high-quality proprietary data are available. This approach is particularly advantageous in scenarios where low latency and enhanced privacy are critical, such as in gaming applications and smart portable devices. By processing data locally on the device, on-device deployments reduce the need for data transmission to external servers, thereby minimizing latency and protecting user privacy.

Technical complexity and data governance also shape deployment choices. Cloud solutions offer user-friendly tools and immediate access to advanced models, benefiting organizations with limited technical resources. However, organizations handling sensitive data or requiring strict compliance might prefer on-premises or on-device deployment despite its technical demands.

**Lessons Learned from Past Implementation**

Success in LLM adoption requires balancing technical capabilities with business objectives. Organizations should begin by identifying specific, high-value use cases where LLMs can deliver measurable impact. This focused approach enables better resource allocation and helps demonstrate early successes, building momentum for broader adoption. The initial use cases should address clear business needs while remaining technically feasible within the organization's current capabilities. Embarking on a small pilot project allows you to demonstrate value, build momentum, and scale to larger use cases. This approach enables quick wins, fosters team motivation, and facilitates gradual scaling, ensuring sustainable success.

It is also important to start simple without big resource commitments. Try prompt engineering with a few examples. Add RAG to address factual accuracy and frequent data updates. If all simple trials do not work, you can fine-tune with carefully designed high quality data.

Building internal expertise is crucial while maintaining strategic partnerships with external providers. Organizations should invest in training technical teams on LLM fundamentals, prompt engineering, RAG, fine-tuning, and application deployment best practices such as data and model version controls, monitoring, alerts, and re-training. Simultaneously, partnerships with experienced vendors can accelerate implementation and provide access to specialized knowledge. This dual approach helps organizations develop self-sufficiency while leveraging external expertise for complex challenges.

Security and monitoring frameworks must be established from the outset. This includes implementing robust data protection measures, regular security audits, and continuous monitoring of model performance and usage patterns. Organizations should develop clear protocols for data handling, model access controls, and incident response procedures. Regular security assessments help identify and address potential vulnerabilities before they become critical issues.

Establishing clear performance metrics and feedback mechanisms ensures continuous improvement. Organizations should define both offline and online A/B testing metrics aligned with business objectives. These metrics should encompass technical performance indicators like response accuracy and latency, as well as business impact measures such as user satisfaction and process efficiency. Regular feedback collection from end-users and stakeholders helps refine the implementation and guides future developments. A data-driven culture equipped with an experimentation platform is critical for success.

The path to successful LLM adoption requires ongoing commitment to governance and refinement. Regular reviews of implementation plans and achievement of key milestones, key evaluation metrics and measurements, and security measures ensure the technology continues to serve business needs effectively while managing associated risks.

**Conclusion**

As LLM technology continues to evolve, this framework provides a foundational structure for future developments and deployments. It enables organizations to systematically evaluate and implement LLM solutions while maintaining flexibility for emerging capabilities and use cases. With "model-as-a-service" based on foundational base models, the importance of data and evaluation becomes increasingly pronounced [16]. In business applications, competitive advantage is shifting from model architecture design and training to high-quality data, model adaptations, and evaluation and measurement, especially in practical settings for typical enterprises. By providing this structured approach, the paper contributes to both academic understanding and practical implementation of LLM technologies, fostering innovation and effective utilization across diverse business contexts.